\documentclass[twocolumn,showpacs,preprintnumbers,amsmath,amssymb]{revtex4}
\usepackage{graphicx}
\usepackage{dcolumn}
\usepackage{bm}

\newcommand{\ket}[1]{|#1\rangle}

\begin{document}
\preprint{CSPIN/0501}

\title{Inelastic transport in molecular spin valves}% Force line breaks with \\

\author{N.~Jean}
%\altaffiliation[Also at ]{Physics Department, XYZ University.}
\author{S.~Sanvito}%
\email{sanvitos@tcd.ie}
\affiliation{School of Physics, Trinity College, Dublin 2, IRELAND}

\date{\today}

\begin{abstract}
We present a study of the effects of inelastic scattering on the transport properties of
various nanoscale devices, namely H$_2$ molecules sandwiched between Pt contacts, 
and a spin-valve made by an organic molecule attached to model half-metal ferromagnetic 
current/voltage probes. In both cases we use a tight-binding Su-Schrieffer-Heeger
Hamiltonian and the inelastic effects are treated with a multi-channel method, including 
Pauli exclusion principle.
In the case of the H$_2$ molecule, we find that inelastic backscattering is responsible for
the drop of the differential conductance at biases larger than 
the excitation energy of the lower of the molecular phonon modes.
In the case of the spin-valve, we investigate the different spin-currents and the 
magnetoresistance as a function of the position of the Fermi level with respect 
to the spin-polarized band edges. In general inelastic scattering reduces the
spin-polarization of the current and consequently the magnetoresistance.
\end{abstract}
\maketitle
\section{Introduction}
Recent years have witnessed an increasing interest for spintronics
devices and for their possible application as novel elements in
microelectronics industry \cite{spintronics,spintronics5,spintronics6}. 
Read-heads based on the giant magnetoresistance (GMR) effect \cite{GMR1,GMR2}
have already revolutionized the hard-drives market and other magnetic devices
such as magnetic random access memory (MRAM) are expected to have a similar 
or even bigger impact. These devices exploit both the magnetic and the electronic 
properties of conducting materials. 

In the typical spin-valve geometry a
non-magnetic spacer is sandwiched between two magnetic contacts. These provide
a source of spin-polarized electrons and the device practically behaves as
a polarizer/analyzer spin filter, with the magnetization vector establishing
the polarization axis. Typically the resistance of the device is higher when
the magnetization vectors of the contacts are aligned antiparallel to each
other, and drops when the alignment is parallel.
The non-magnetic spacer controls the total resistance of the device
and the overall transport regime (diffusive, ballistic, tunneling, etc.),
while the quality and nature of the interfaces can further affect the spin 
polarization.

Many suggestions for possible non-magnetic spacers have been recently
proposed \cite{spintronics7}. Among the most intriguing there is the
possibility of using organic molecules \cite{Alex}. These offer low
fabrication costs and high manufacturability typical of molecular
electronics \cite{Joanat,Rat_rectifiers} and the advantage of long
spin-relaxation times. Several prototypes for molecular spin-valves have
been already fabricated and GMR in excess of 10\% have been reported
for spacers such as $\pi$-conjugated molecules \cite{Shi,Dediu}
and in organic tunneling junctions \cite{Ralph}. 

Electron conduction through these systems is always explained in terms of elastic 
transport \cite{Alex,pati,kirk} and inelastic effects are often neglected. 
This may be a quite drastic simplification in particular in the case of thick
organic spacers, for which the transmission due to direct tunneling becomes 
negligible. Inelastic scattering not only does open new transport channels but
also affects the transmission of the elastic ones by introducing competition for 
the same final states due to the Pauli exclusion principle \cite{Datta,Eldon}.
These effects can change drastically the conducting properties of the
devices and affect the GMR of spin-valves. Recent experimental and theoretical 
works have already shown the importance of the inelastic scattering in different
non-magnetic molecular devices
\cite{Ness,Ratner,Magoga,Sautet,Agrait,Fred,spintronics3,Montgomery}.

In this work we investigate the effects of inelastic scattering on the
transport properties of model spin-valves. First, we briefly introduce 
our method for calculating the $I$-$V$ characteristics of a typical two probe
device in presence of electron-phonon coupling. This is based on a combination
of scattering theory in a Green's function framework \cite{Stefano1,Stefano2}
and a self-consistent evaluation of the non-equilibrium electron distribution 
\cite{Eldon}. Then we show a typical calculation by investigating the
transport properties of the H$_2$ molecule sandwiched between two platinum leads.
Finally we consider the case of organic spin-valves obtained by
sandwiching a mono-atomic chain, resembling the properties of a (CH)$_n$ 
molecule, between half-metallic leads. For this system
we consider a model $M_{\beta^\prime}$ half metal (according to the classification
of Coey and Sanvito \cite{HM}), which resembles the electronic structure of magnetite.

%%%%%%%%%%%%%%%%%%%%%%%%%%%%%%%%%%%%%%%%%

\section{Computational Method}
%
%%%%%%%%%%%%%%%%%%%%%%%%%%%%%%%%%%%%%%%%%
%
\subsection{Problem set-up}
\begin{figure}[htbp]
\centering
\includegraphics[height=0.6truecm,width=7truecm]{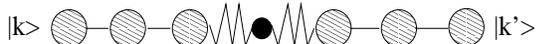}
\caption{{Schematic representation of a system composed
of two semi-infinite current/voltage leads and a scattering region, where inelastic
scattering can occur. An electron with $k$-vector $k$ approaching the scattering region from 
the left is transmitted to the right with $k$-vector $k^\prime$. In this 1D case if
$k\ne k^\prime$ the energy is not conserved.}}
\label{Fig1}
\end{figure}

The system under investigation is schematically represented in
figure \ref{Fig1}. It consists of two semi-infinite current/voltage probes 
formed from mono-atomic chains and sandwiching a scattering region
(the molecule).  Electron-phonon interaction is present only in the scattering region
while the leads act as an ideal phonon sink.
The scattering region is composed by one or more atoms connected to the
leads both via electron and phonon-assisted hopping. 

The Hamiltonian $H$ describing such a system can be divided in five parts
\begin{equation}
H=H_\mathrm{L}+V_\mathrm{ML}+H_\mathrm{M}+V_\mathrm{MR}+H_\mathrm{R}\:.
\label{hamil}
\end{equation}
$H_\mathrm{L}$ ($H_\mathrm{R}$) describes the left-hand side (right-hand side)
lead, $H_\mathrm{M}$ describes the molecule and $V_\mathrm{ML}$ ($V_\mathrm{MR}$)
the interaction between the left (right) lead and the molecule.
For the leads we consider a tight-binding model expanded over a $s$-like
orthogonal basis set $\ket{i}$, where $i$ labels the atomic position. Although there is
no conceptual complication in considering a multi-orbital tight-binding model here we use 
only one orbital per atomic site in order to keep the computational overheads to a minimum. 
This choice is motivated by the severe scaling of our computational method \cite{Eldon} 
with the number of phonons. Thus $H_\mathrm{L}$ and $H_\mathrm{R}$ read
\begin{eqnarray}
H_\mathrm{L}=\epsilon_\mathrm{L} \sum_{i=-\infty}^{0}c_{i}^{\dag} c_{i} + \beta_\mathrm{L}
\sum_{i=-\infty}^{-1}(c_{i}^{\dag} c_{i+1}+ c.c.)\:, \\
H_\mathrm{R}=\epsilon_\mathrm{R} \sum_{i=N+1}^{+\infty} c_{i}^{\dag} c_{i} + \beta_\mathrm{R}
\sum_{i=N+1}^{+\infty}(c_{i}^{\dag} c_{i+1}+ c.c.)\:.
\label{HL}
\end{eqnarray}
In the expressions above $c_i$ and $c_i^{\dag}$ are the annihilation and creation operators
for electrons at the site $i$, $\beta_\alpha$ ($\alpha$= R, L) is the hopping parameter in the leads 
and $\epsilon_\alpha$ is the lead on-site energy. This already include the rigid shift given 
by an external voltage $V$ applied to the system, that is 
$\epsilon_\mathrm{L}=\epsilon_\mathrm{L}^0 +\frac{V}{2}$ and
$\epsilon_\mathrm{R}=\epsilon_\mathrm{R}^0 -\frac{V}{2}$, with 
$\epsilon_\mathrm{L/R}^0$ the on-site energy for $V=0$. Note that we implicitly assume
that the leads are good metals presenting local charge neutrality \cite{Alex2}. 
The leads extend from $- \infty$ to $i=0$ to the left and from 
$N+1$ to $+\infty$ to the right, with the molecule comprising $N$ sites ($i=1, N$).

\begin{figure}[htbp]
\centering
\includegraphics[height=2.0truecm,width=9truecm]{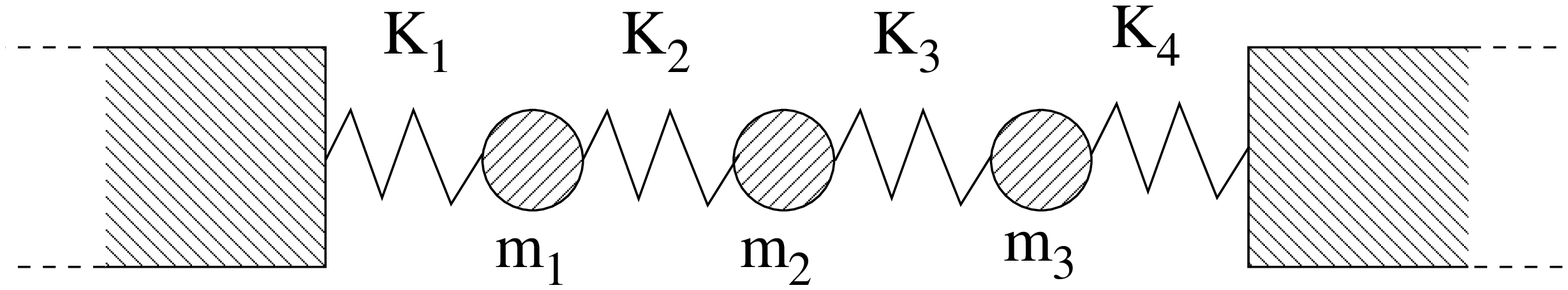}
\caption{{Classical ``springs and balls'' model used to calculate the phonon modes. 
The leads act as hard walls constraining the
oscillation to the middle atomic chain. $m_j$ are the masses of the
atoms in the scattering region and $K_j$ are the spring constants.}}
\label{mapping3}
\end{figure}

In our model system the electron-phonon interaction is present only in the
molecule and at the interface of this with the leads. We use for the description of 
such a scattering region the Su-Schrieffer-Heeger (SSH) Hamiltonian 
\cite{SSH,SSH1,SSH2} in its second quantization form, considering only
longitudinal phonon modes \cite{Montgomery}. Thus $H_\mathrm{M}$ reads
\begin{eqnarray}
H_\mathrm{M}=\epsilon_\mathrm{M} \sum_{i=1}^{N} b_{i}^{\dag} b_{i} + \sum_{i \neq j }
t_{ij}(b_{i}^{\dag} b_{j}+b_{j}^{\dag} b_{i}) + \sum_k \hbar \omega_k a_k^{\dag} a_k +\nonumber \\
+\sum_{k}\sum_{i \neq j}
\gamma_{ij}^{k}(a_k^{\dag}+a_k)(b_{i}^{\dag} b_{j}+b_{j}^{\dag}
b_{i}) \;.
\label{MOL}
\end{eqnarray}
Such a SSH-like Hamiltonian describes linear electron-phonon coupling. It has a purely electronic 
part written in a tight-binding form with annihilation and creation operators $b_{i}$ and 
$b_{i}^{\dag}$, hopping parameter $t_{ij}$ and on-site energy $\epsilon_\mathrm{M}$. 
These tight-binding parameters refer to the static system (without phonons). In addition
there is an electron-phonon interaction part $\sum_{k}\sum_{i \neq j}
\gamma_{ij}^{k}(a_k^{\dag}+a_k)(b_{i}^{\dag} b_{j}+b_{j}^{\dag}
b_{i})$, which describes phonon absorption/emission. This is linear in
the phonon annihilation and creation operators $a_k$ and $a_k^{\dag}$ with the coupling strength 
given by $\gamma_{ij}^{k}$. Finally it comprises the purely phononic energy 
$\sum_k \hbar \omega_k a_k^{\dag} a_k$ with $\omega_k$ the phonon frequency
for the $k$-mode. The frequencies $\omega_k$ are calculated using a classical
``springs and balls'' model for the scattering region \cite{Kittel}, i.e. by diagonalizing the
dynamical matrix obtained with the appropriate boundary conditions (see figure \ref{mapping3}).

The parameter $\gamma_{ij}^{k}$ depends on both the energy of
the $k$-mode, $\hbar\omega_k$, and the vibrational properties of the atomic sites
$i$ and $j$. It can be written as
\begin{eqnarray}
\gamma_{ij}^k=\alpha_{ij}\left\{\frac{e_k^i}{\sqrt{m_i}} -
\frac{e_k^j}{\sqrt{m_j}}\right\} \sqrt{\frac{\hbar}{2 \omega_k}}
\label{gammahop}
\end{eqnarray}
where $e_k^i$ is the displacement vector of the $i$-th atom associated to the $k$-mode, $m_{i}$
($m_j$) is the mass of the $i$-th ($j$-th) atom and $\alpha_{ij}$ is the coefficient of
the linear expansion of the hopping parameter over the atomic
displacement
\begin{eqnarray}
t(q_i-q_j)=t_{ij}+\alpha_{ij}(q_i-q_j)\:.
\end{eqnarray}
In the previous equation $t_{ij}$ is the first term of the expansion and stands for the
hopping parameter of the static system at the equilibrium distance, while $q_i$ and
$q_j$ are the displacements from the equilibrium positions of the atoms $i$ and $j$
due to the lattice vibrations. 

The Hamiltonian coupling the molecule to the leads is again of a SSH form
\begin{eqnarray}
V_\mathrm{ML}=\left[w^0+\sum_k\Gamma^k(a_k^{\dag}+a_k)\right](c_0^{\dag}b_1 +b_1^{\dag}c_0)\:,
\end{eqnarray}
with an analogous expression for $V_\mathrm{MR}$.
The purely elastic part is described by the hopping
parameter $w^{0}$ while the inelastic part is characterized by the
electron phonon coupling $\Gamma^k$
\begin{eqnarray}
\Gamma^k=\alpha_{1}\frac{e_k^1}{\sqrt{m_1}} \sqrt{\frac{\hbar}{2 \omega_k}}\;.
\label{gammahop2}
\end{eqnarray}
The equation (\ref{gammahop2}) is obtained from the (\ref{gammahop}) by
imposing static boundary conditions to the leads ($q_i=0$ for $i$ belonging to the leads). 
 
\subsection{Inelastic scattering as an effective elastic problem}

We tackle the problem of transport in presence of electron-phonon interaction
by using the method proposed first by Bon$\check{\mathrm{c}}$a and Trugman \cite{BoncaTrug,HauleBonca}. 
This approach consists in mapping a general many-body transport problem onto a fictitious 
one-body problem, where only elastic scattering is present. 
The method, in contrast to other perturbative approaches \cite{SBA}, is in principle exact. i.e. 
it does not introduce any approximation in calculating the inelastic contributions to the current.
The drawback is cast in the dimension of the Hilbert space needed for defining the fictitious
one-body system, which grows rapidly with the number of phonon states considered.
In this section we will restrain our attention to mono-atomic single-channel leads,
but the same procedure can be easily extended to the multichannel case \cite{Nicola2} 

The main idea behind this approach is that the energy is globally conserved
during the scattering process. In fact, although an electron leaving the 
left-hand side lead with energy $E$ and emitting a phonon is absorbed by the 
right-hand side lead with a different energy $E^\prime$, the energy of the
system electron plus phonon is conserved ($E-E^\prime$ is the
phonon energy). This suggests the possibility of constructing an associated multi-channel scattering 
problem \cite{LBP}, using asymptotic states that include both electronic and phononic
degrees of freedom, all having the same total energy. In this way the scattering problem is
formally elastic and can be solved by standard scattering theory \cite{Datta,Stefano1}.
However, when extracting the electron current, one should be able to map this fictitious problem 
back to the original many-body one.

We then define states $\ket{k ; n_\alpha, n_\beta,...}$ where the electronic part 
$\ket{k}$ describes an asymptotic Bloch state $k$ in the leads, and the
phononic part $\ket{n_\alpha, n_\beta,...}$ is a Fock state associated
to the various phonon modes $\alpha$, $\beta$, ... of the molecule. Here $n_\alpha$
is the occupation number of the mode $\alpha$. 
During a generic scattering event the total energy of such
state $E=E_k+\hbar\sum_\alpha(n_\alpha\omega_\alpha)$ is conserved, despite the fact that
the electronic energy $E_k$ may not be. Therefore the scattering between the asymptotic states
$\ket{k ; n_\alpha, n_\beta,...}$ of this fictitious system is formally elastic.
\begin{figure}[htbp]
\centering
\includegraphics[height=6.5truecm,width=8truecm]{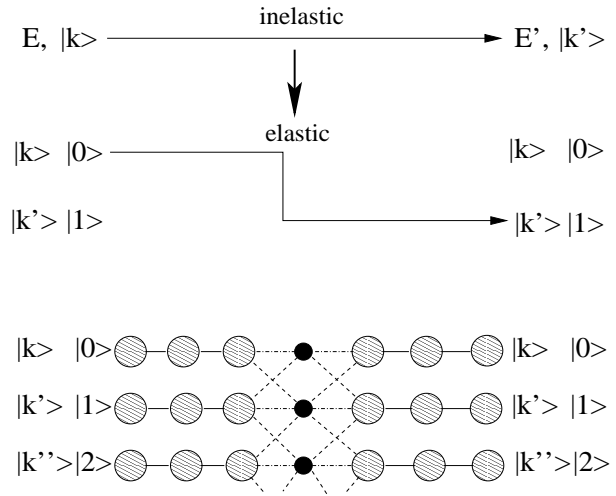}
\caption{{Mapping procedure for a mono-atomic system with
only one phonon mode. (a) During an inelastic process electrons change their
energy from $E$ to $E^\prime$. In contrast in our fictitious system the
scattering between states with different electronic energy 
($\ket{k,0}\rightarrow\ket{k^\prime,1}$) is elastic,
since the total energy (electron plus phonon) is conserved.
(b) A mono-atomic chain sandwiching a molecule with vibrational degrees of freedom
is mapped onto an equivalent system of disconnected atomic chains coupled through 
a scattering region.}}
\label{mapping4}
\end{figure}
Note however that the Bloch states $\ket{k}$ associated to the different asymptotic states 
$\ket{k ; n_\alpha, n_\beta,...}$ in general have different energies, and must be calculated
from the electronic secular equation at the energy 
$E_k=E-\hbar\sum_\alpha(n_\alpha\omega_\alpha)$. Importantly if the scattering is
between states having identical phononic parts, then it is elastic not only for the 
fictitious system, but also for the actual many-body problem.

Therefore the mapping $\ket{k}\rightarrow\ket{k ; n_\alpha, n_\beta,...}$ transforms
an inelastic problem onto a fictitious multi-channel elastic one (see figure \ref{mapping4}). 
For example within this approach a  linear atomic chain sandwiching a vibrating molecule
is formally equivalent to a collection of linear chains sandwiching an elastic scatterer,
where the coupling between the chains is provided by the electron-phonon interaction
only (see figure \ref{mapping4}).

One can now construct a scattering theory for the asymptotic states $\ket{k ; n_\alpha, n_\beta,...}$ 
by using one of the several methods available for elastic transport \cite{Stefano2}.
In this case however there are two additional features. First a new expression for 
the current, including Pauli exclusion principle, should be derived. In fact different 
incoming asymptotic states corresponding to electronic states with different energy 
may compete for the same final
Bloch state. Secondly the dimension of the Hilbert space is in principle infinite, since
the occupation numbers entering the states $\ket{k ; n_\alpha, n_\beta,...}$ are unbound. 
In practice however only a finite number of
phonon modes is usually sufficient to obtain converged results, that is when the scattering
transmission coefficients do not vary with the number of phonon states.
Most importantly the convergence can be carefully monitored by progressively enlarging 
the Hilbert space.

\subsection{Scattering theory}

In the fictitious system, where the asymptotic states are $\ket{k ; n_\alpha, n_\beta,...}$,
the scattering is formally only elastic. Therefore we can calculate the 
quantum mechanical scattering probabilities by using standard scattering theory.
In this work we have used the Green's function approach introduced by
Sanvito et al. \cite{Stefano1}. The method consists in evaluating the total
retarded Green's function for the entire system (leads plus scattering region),
from which extracting the scattering matrix. Here we briefly summarize the 
method and highlight the main modifications needed when dealing with the
fictitious system, instead of a purely electronic one. Furthermore we
consider simple one-dimensional leads, supporting only one scattering channel
at a given energy.

The starting point consists in writing the surface Green's function for the leads. 
These are in the form of a block diagonal matrix, which, in the case of the
left-hand side lead, reads
\begin{displaymath}
G^0_\mathrm{L}(E)=
\left(\begin{array}{cccc}
{g}_\mathrm{L}^0(E) & 0 & \ldots & \ldots \\
0 & g_\mathrm{L}^\mu(E^\prime) & \ldots & \ldots \\
\ldots &\ldots  &  g_\mathrm{L}^{\mu^\prime}(E^{\prime\prime})& \ldots \\
\ldots &\ldots  & \ldots & \ldots \\
\end{array} \right)\;.
\label{matriciona}
\end{displaymath}
An analogous expression is valid for the right-hand side lead surface Green's function
$G^0_\mathrm{R}(E)$.
${g}_\mathrm{L}^\mu(E^\prime)$ is the left surface Green's function of the 
purely electronic system (no phonons) at the energy $E^\prime$. In general it is a 
$N\times N$ matrix, with $N$ the number of degrees of freedom (orbitals) describing
the leads surface, but in this simple case it is only a c-number
\begin{equation}
{g}_\mathrm{L}^\mu(E^\prime)=\frac{\mathrm{e}^{ik}}{\beta_\mathrm{L}}\:,
\end{equation}
with $k$ the adimensional $k$-vector solution of the secular equation
for the simple chain
\begin{equation}
E^\prime=\epsilon_\mathrm{L}+2\beta_\mathrm{L}\cos(k)\:.
\end{equation}

$\mu$ is a collective
index, which label the phononic state $\mu=\{n_\alpha,n_\beta,...\}$ and the energy
$E^\prime$ is the electronic energy of the state $\ket{k ; n_\alpha, n_\beta,...}$,
i.e. $E^\prime=E-\hbar\sum_\alpha(n_\alpha\omega_\alpha)$.
Note that the block diagonal form of $G_\mathrm{L}^0$ and $G_\mathrm{R}^0$
is the result of the fact that different phonon states $\ket{n_\alpha, n_\beta, ..}$
do not mix in the leads. 
Note also that the Green's functions $G_\mathrm{L}^0$ and $G_\mathrm{R}^0$ 
are in principle infinite 
matrices since the number of possible phonon states is infinite. In practice states 
with very large occupation numbers are not accessible and a cutoff $N_\mathrm{ph}$
over the number of phonon states is used.

In contrast to what happens in the leads, different phononic states
do interact inside the scattering region. The matrix representing
such an interaction is a $(M\cdot N_\mathrm{ph})\times(M\cdot N_\mathrm{ph})$
hermitian matrix, where $M$ is the number of electronic degrees of
freedom (orbitals) describing the molecule. However, it is important to note
that in this simple scattering approach, where no self-consistent evaluation
of the scattering potential is carried out, it is possible to eliminate the
explicit dependence of the scattering region Hamiltonian over the 
internal electronic degrees of freedom. This is done by using the 
recursive decimation method \cite{Stefano1}, to yield an effective
energy-dependent $2N_\mathrm{ph}\times 2N_\mathrm{ph}$ 
Hamiltonian $H_\mathrm{eff}(E)$, coupling all the phononic states
in the left-hand side lead with those of the right-hand side lead
\begin{equation}
H_{\rm{eff}}(E)=\left(
\begin{array}{rr}
H_{\rm{L}}^*(E) & H_{\rm{LR}}^*(E)\\
H_{\rm{RL}}^*(E) & H_{\rm{R}}^*(E)\\
\end{array}
\right){\;}.
\end{equation}
The blocks $H_{\rm{L}}^*$ and $H_{\rm{R}}^*$ describe the interaction 
respectively within the last plane of the left-hand side and first plane of 
the right-hand side lead, while the off diagonal blocks describe the 
coupling between these planes.

Finally the total Green's function of the system leads plus molecule
calculated at the two extremal planes of the leads ($i$=0 and $i$=$N$+1) 
can be obtained by
simply solving Dyson's equation
\begin{equation}
G(E)=[G^0(E)^{-1}-H_\mathrm{eff}(E)]^{-1}=
\left(\begin{array}{cc}
G_\mathrm{L} & G_\mathrm{LR} \\
G_\mathrm{RL} & G_\mathrm{R}\\
\end{array}
\right)\;,
\end{equation}
where $G^0(E)$ is written as
\begin{equation}
G^0(E)=\left(
\begin{array}{cc}
G_\mathrm{L}^0(E) & 0 \\
0 & G_\mathrm{R}^0(E)\\
\end{array}
\right){\;}.
\end{equation}

The total Green's function $G$ contains all the information about the
scattering amplitudes. These can be explicitly extracted by using the 
Fisher and Lee relation \cite{FisherLee} or in more general terms
by projecting $G$ over a general scattering wave-function \cite{Stefano1,
Stefano2}. For our single-mode leads these simply read

\begin{equation}
t_{\mu\nu}=i\hbar(G_\mathrm{LR})_{\mu\nu}\sqrt{v_\mu v_\nu}\;,
\end{equation}
\begin{equation}
r_{\mu\nu}=-\delta_{\mu\nu}+i\hbar(G_\mathrm{L})_{\mu\nu}\sqrt{v_\mu v_\nu}\;,
\end{equation}
\begin{equation}
t_{\mu\nu}^{\prime}=i\hbar(G_\mathrm{RL})_{\mu\nu}\sqrt{v_\mu v_\nu}\;,
\end{equation}
\begin{equation}
r_{\mu\nu}^{\prime}=-\delta_{\mu\nu}+i\hbar(G_\mathrm{R})_{\mu\nu}\sqrt{v_\mu v_\nu}\;,
\end{equation}

where $v_{\mu}$ is the group velocity associated to the electronic component of the
state $\ket{\mu}=\ket{k ; n_\alpha, n_\beta,...}$
\begin{equation}
v_{\mu}=-\frac{2\beta_\mathrm{L}}{\hbar}\sin(k)\;.
\end{equation}

$t_{\mu\nu}$ and $r_{\mu\nu}$ are respectively the transmission
and reflection amplitude for states incoming from the left-hand side lead,
while $t_{\mu\nu}^{\prime}$ and $r_{\mu\nu}^{\prime}$ are the same
quantities but for states incoming from the right-hand side lead. 

Finally the current can be extracted from the Landauer formula \cite{Datta}
\begin{equation}
I=\frac{2 e}{h}  \int \sum_{\mu\nu}[|t_{\mu\nu}|^2F_L(E)-
|t_{\mu\nu}^{\prime}|^2 F_R(E)]dE\;,
\label{curnopauli}
\end{equation}
where $F_L(E)$ and $F_R(E)$ are the Fermi distributions respectively for 
left- and right-hand side lead. Note that in the case of time-reversal symmetry
$t_{\mu\nu}=t_{\mu\nu}^\prime$ and the current becomes
\begin{equation}
I=\frac{2 e}{h}  \int \sum_{\mu\nu}|t_{\mu\nu}(E)|^2[F_L(E)-F_R(E)]dE\;.
\label{curnopauli2}
\end{equation}

%%%%%%%%%%%%%%%%%%%%%%%%%%%%%

\subsection{Pauli principle} 

For elastic transport different scattering channels do not
compete for the same final state, since the scattering wave-functions
extend from one lead to the other. This is also formally true for our
fictitious system. However when one considers the electronic part
of the fictitious system, the situation appears rather different. Consider 
for example the scattering process in which $\ket{k ; n_\alpha, n_\beta,...}$
is transmitted into the same state $\ket{k ; n_\alpha, n_\beta,...}$, and the
one in which it is the state $\ket{k^\prime ; m_\alpha, m_\beta,...}$
to be transmitted into $\ket{k ; n_\alpha, n_\beta,...}$.
In the first process the phononic configurations of the initial
and final states are identical and therefore the scattering is elastic. 
In contrast the second process involves a change in phononic configuration, 
which means that energy has been exchanged between the electronic and the
phononic sub-systems. This is an inelastic scattering event.

The crucial point is that these two processes lead to the same final
electronic state $\ket{k}$, the first through elastic and the 
second through inelastic scattering. These two scattering events
do not share the same scattering wave-function and therefore compete
for the same electronic states. Therefore in this case the Pauli
exclusion principle should be taken explicitly into consideration
when evaluating the current. 

We then use the procedure introduced by Emberly and Kirczenow 
\cite{Eldon}, which evaluates self-consistently the non-equilibrium electron 
distributions resulting from the Pauli principle. We define 
$(f_{+,el}^{\alpha \alpha})^R(E^\prime)$ as the
electronic distribution of the right-hand side lead resulting 
from electrons transmitted elastically from left to right
at an energy $E^\prime$. As a matter of notation we label with ``$+$''
(``$-$'') scattering processes involving electron transmission
(reflection), ``R'' (``L'') indicates
that the distribution is for the right-hand side (left-hand side)
lead, and {\it ``el''} that the scattering process is elastic. Finally
the indexes $\alpha$ label the associated states
$\ket{\alpha}=\ket{k ; n_\alpha, n_\beta,...}$ describing the process.
\begin{figure}[htbp]
\centering
\includegraphics[height=5.3truecm,width=8.5truecm]{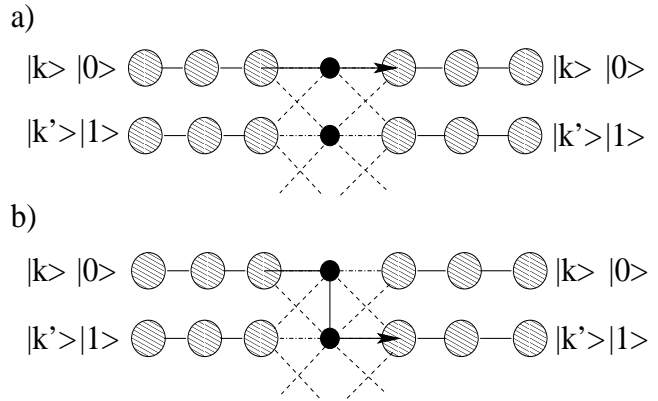}
\caption{{Scattering events of the fictitious system,
contributing to the various electronic distributions. (a)
elastic scattering of the state $\ket{k;0}=\ket{k}\ket{0}$,
contributing to the distribution $( f_{+,el}^{\alpha \alpha} )^R(E^\prime)$. 
(b) inelastic scattering of the state $\ket{k;0}=\ket{k}\ket{0}$ into
the state $\ket{k^\prime;1}=\ket{k^\prime}\ket{1}$
contributing to the distribution $(f_{+,in}^{\alpha \alpha^\prime})^R(E^\prime)$.}}
\label{mapping1}
\end{figure}
Similarly $(f_{+,in}^{\alpha \alpha^\prime})^R(E^\prime)$ is the electronic
distributions in the right-hand side lead given by electrons 
incoming from the left-hand side lead and inelastically transmitted through 
the molecule (Fig \ref{mapping1}). In this case $E^\prime$ is the 
energy of the final electronic state and the scattering is between the 
states $\ket{\alpha}=\ket{k ; n_\alpha, n_\beta,...}$ and
$\ket{\alpha^\prime}=\ket{k^\prime ; m_\alpha, m_\beta,...}$.

If one neglects Pauli's principle these quantities are simply
related to the Fermi distributions of the lead containing the
incoming electrons
\begin{equation}
(f_{+,el}^{\alpha \alpha})^R(E^\prime)=F_L(E^\prime) T^{\alpha \alpha}
(E^\prime,E^\prime)\;,
\end{equation}
\begin{equation}
(f_{+,in}^{\alpha \alpha^\prime} )^R(E^\prime)=F_L(E) 
T^{\alpha \alpha^\prime}(E,E^\prime)\;,
\end{equation}
where we have introduced the transmission coefficients 
$T^{\alpha\beta}(E,E^\prime)=|t_{\alpha\beta}|^2$, which depend
on both the initial $E$ and final $E^\prime$ electronic energy.
All the other distributions, for electrons approaching from the
right-hand side lead, can be written in a completely analogous way
\cite{Eldon}.

We now need to establish a procedure for evaluating the distributions
$f$ in such a way of accounting for Pauli exclusion principle. This is easily done
by imposing that a generic distribution $f$ is given by the combined 
probability of transmitting an electron with that of finding the 
final state empty. For instance the distribution of elastically
transmitted electrons in the right-hand side lead reads
\begin{eqnarray}
(f_{+,el}^{\alpha \alpha})^R(E^\prime)=F_L(E^\prime)c(E^\prime)
T^{\alpha\alpha}(E^\prime,E^\prime)\times \nonumber \\
\bigl[1-\sum_{\alpha^\prime \neq \alpha}(f_{+,in}^{\alpha \alpha^\prime})^R
(E^\prime)-\sum_{\alpha^\prime\neq\alpha} (f_{-,in}^{\alpha \alpha^\prime})^R
(E^\prime) \bigr]\;.
\label{form1} 
\end{eqnarray}
This means that the distribution in the right-hand side lead of electrons
elastically transmitted at an energy $E^\prime$ is proportional to the
probability of transmitting an initially filled state in the left-hand 
side lead $F_L(E^\prime)T^{\alpha\alpha}(E^\prime,E^\prime)$, 
and to the probability that the final state is not already occupied
$\bigl[1-\sum_{\alpha^\prime \neq \alpha}(f_{+,in}^{\alpha \alpha^\prime})^R
(E^\prime)-\sum_{\alpha^\prime\neq\alpha} (f_{-,in}^{\alpha \alpha^\prime})^R
(E^\prime) \bigr]$. This last probability is determined by two
competing effects: inelastic transmission from the left-hand side lead 
$(f_{+,in}^{\alpha \alpha^\prime})^R$, and inelastic reflection 
from the right-hand side lead $(f_{-,in}^{\alpha \alpha^\prime})^R$.

Similar equations can be derived for all the other distributions
in the right-hand side lead
\begin{eqnarray}
(f_{+,in}^{\alpha \alpha^{\prime}})^R(E^\prime)=F_L(E)c(E)T^{\alpha
\alpha^\prime}(E,E^\prime) \times \nonumber \\ 
\bigl[1-(f_{+,el}^{\alpha\alpha})^R(E^\prime)-(f_{-,el}^{\alpha \alpha})^R(E^\prime)- \nonumber \\ 
\sum_{\beta\neq \alpha}(f_{+,in}^{\alpha\beta})^R(E^\prime)-
\sum_{\alpha^\prime\neq \alpha}(f_{-,in}^{\alpha\alpha^\prime})^R(E) \bigr]\;,
\label{form2} 
\end{eqnarray}
\begin{eqnarray}
(f_{-,el}^{\alpha \alpha})^R(E^\prime) = F_R(E^\prime)d(E^\prime) R^{\alpha \alpha}(E^\prime,E^\prime) \times \nonumber \\ 
\bigl[1-\sum_{\alpha^\prime\neq\alpha}(f_{+,in}^{\alpha\alpha^\prime})^R(E^\prime)-
\sum_{\alpha^\prime\neq\alpha}(f_{-,in}^{\alpha \alpha^\prime})^R(E^\prime) \bigr]\;,
\label{form3} 
\end{eqnarray}
\begin{eqnarray}
(f_{-,in}^{\alpha \alpha^{\prime}})^R(E^\prime)=F_R(E)d(E)R^{\alpha\alpha^\prime}
(E,E^\prime) \times \nonumber \\ 
\bigl[1-(f_{+,el}^{\alpha\alpha})^R(E^\prime)-(f_{-,el}^{\alpha\alpha})^R(E^\prime)-\nonumber \\ 
\sum_{\beta\neq\alpha}(f_{-,in}^{\alpha\beta})^R(E^\prime)-
\sum_{\alpha^\prime\neq\alpha}(f_{+,in}^{\alpha\alpha^\prime})^R(E) \bigr]\;,
\label{form4} 
\end{eqnarray}
where $(f_{+,in}^{\alpha \alpha^{\prime}})^R(E^\prime)$, 
$(f_{-,el}^{\alpha \alpha})^R(E^\prime)$ and 
$(f_{-,in}^{\alpha \alpha^{\prime}})^R(E^\prime)$ are
the distributions for electron undergoing respectively to inelastic 
transmission, elastic reflection and inelastic reflection, and we
have introduced the reflection coefficient
$R^{\alpha\beta}(E,E^\prime)=|r_{\alpha\beta}|^2$. Analogous
expression for the left-hand side lead can be obtained 
by replacing $R\rightarrow L$, $T^{\alpha\beta}\rightarrow 
T^{\alpha\beta\prime}=|t_{\alpha\beta}^\prime|^2$ and
$R^{\alpha\beta}\rightarrow 
R^{\alpha\beta\prime}=|r_{\alpha\beta}^\prime|^2$ in the equations
(\ref{form1}) through (\ref{form4}).

The normalization functions $c(E)$ and $d(E)$ are obtained by imposing 
charge conservation deep into the leads, which means that the sum of all the 
distributions $f$ must be equal to the equilibrium Fermi distribution
of the particular lead
\begin{eqnarray}
F_L(E)=(f_{+,el}^{\alpha \alpha})^R(E)+(f_{-,el}^{\alpha\alpha})^L(E)+\nonumber \\ 
\sum_{\alpha^\prime}(f_{+,in}^{\alpha \alpha^\prime})^R(E^\prime)
+\sum_{\alpha^\prime}(f_{-,in}^{\alpha \alpha^\prime})^L(E^\prime)\;,
\label{cons1}
\end{eqnarray}
\begin{eqnarray}
F_R(E) = (f_{+,el}^{\alpha \alpha})^L(E)+(f_{-,el}^{\alpha \alpha})^R(E)+\nonumber \\ 
\sum_{\alpha^\prime}(f_{+,in}^{\alpha \alpha^\prime})^L(E^\prime)
+\sum_{\alpha^\prime}(f_{-,in}^{\alpha \alpha^\prime})^R(E^\prime)\;.
\label{cons2}
\end{eqnarray}
The equations (\ref{cons1}) and (\ref{cons2}) together with those for the
distributions (\ref{form1}), (\ref{form2}), (\ref{form3}) and (\ref{form4})
form a close set of equations, that can be solved for the $f$ iteratively.
These are evaluated numerically over a discrete energy mesh. 
Finally the current is obtained by a direct extension of the Landauer
formula, which takes into account the self-consistent electron
distributions
\begin{eqnarray}
I=\frac{2e}{h}\int\left(\sum_{\alpha}\left[(f_{+,el}^{\alpha\alpha})^R(E)
+\sum_{\alpha^\prime}(f_{+,in}^{\alpha \alpha^\prime})^R(E)\right]-\right. \nonumber \\ 
-\left.\sum_{\alpha}\left[(f_{+,el}^{\alpha\alpha})^L(E)+
\sum_{\alpha^\prime}(f_{+,in}^{\alpha \alpha^\prime})^L(E)\right]\right)dE\;.
\label{currentf}
\end{eqnarray}

%%%%%%%%%%%%%%%%%%%%%%%%%%%%%%%%%%%%%

\section{Results}

We are now in the position of investigating inelastic transport
in atomic scaled systems. In particular we will consider three 
material systems. The first is a simple diatomic molecule. This will be
used as a test case for explaining some general concepts of inelastic
transport and for providing details about the convergence. Then we
will investigate the case of the H$_2$ molecule, for which 
experimental data are available \cite{smit}. Finally we will
move our attention to the investigation of inelastic scattering in
molecular spin-valves made from half-metallic leads. In all the calculations
we consider temperatures considerably lower than the energy of the
lowest of the phonon modes. This means that $\ket{k; 0, 0, ...}$ is
the only occupied incoming channel and that there cannot be a net phonon
absorption.

%%%%%%%%%%%%%%%%%%%%%%%%%%%%%%%%%%%%%%%

\subsection{Fixing the phonon cutoff}

The first system that we consider is a diatomic molecule sandwiched
between two identical mono-dimensional leads. These are described
by a single orbital nearest neighbor tight-binding model with 
on-site energy $\epsilon_\mathrm{L}=\epsilon_\mathrm{R}$=-10~eV and hopping
parameter $\beta_\mathrm{L}=\beta_\mathrm{R}$=2~eV (the total bandwidth is 8~eV).
The Fermi level of the leads is taken at $E_\mathrm{F}$=-10~eV, which 
corresponds to half-filling. 
The tight-binding parameters of the diatomic molecule are
taken from reference \cite{Eldon} and are $\epsilon_M$=-10~eV, $w^0$=0.15~eV,
$t_{ij}$=0.2~eV, $\alpha_{ij}$=6.1~eV/\AA. The masses of the atoms
in the molecule are 13~a.m.u. (each site corresponds to a C-H group) and
the spring constants are $K_\mathrm{lead-M}$=180~eV/\AA$^2$, and 
$K_\mathrm{M}=90$~eV/\AA$^2$, where the first is the one between the 
leads and the molecule and the second is the intra-molecular one.

\begin{figure}[htbp]
\centering
\includegraphics[width=3.5truecm,height=5truecm]{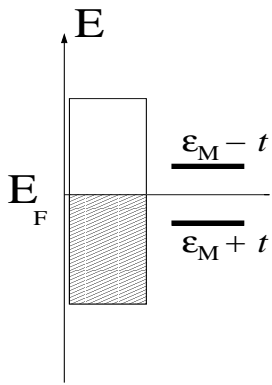}
\caption{{Schematic representation of the energy level lineup
of the diatomic molecule discussed in this section. The solid box represents
the density of states of the mono-atomic leads, the solid horizontal lines 
denote the position of the bonding and antibonding states of the free molecule
and the thin horizontal line is the position of the Fermi level $E_\mathrm{F}$.}}
\label{alli}
\end{figure}

These parameters lead to two longitudinal phonons (the only ones considered
here) of energies $\hbar\omega_1$=0.24~eV and $\hbar\omega_2$=0.33~eV. A
schematic diagram of the energy level alignment is presented in figure
\ref{alli}. The main goal of this section is to discuss how
the energy mesh and the number of phonon states included in the 
calculation affect the results.

The general strategy for testing the convergence of the calculation is 
as follows. First one investigate how the transmission coefficient
changes with the number of phonon states $N_\mathrm{ph}$ included 
in the calculation.
This is done with a reasonably fine energy mesh in order to have 
a good energy resolution. Then, once $N_\mathrm{ph}$ is fixed,
the energy meshed is varied and the convergence of the current is 
controlled. Note that the energy mesh and the cutoff over the
number of phonon states have two deeply different physical meanings.
The energy mesh merely controls how accurate is the energy resolution,
and typically a resolution of the order of 1~meV is sufficient.
In contrast the number of phonon states essentially 
establishes the degree of accuracy in describing multiple phonon
scattering. 

\begin{figure}[htbp]
\centering
\includegraphics[width=9truecm]{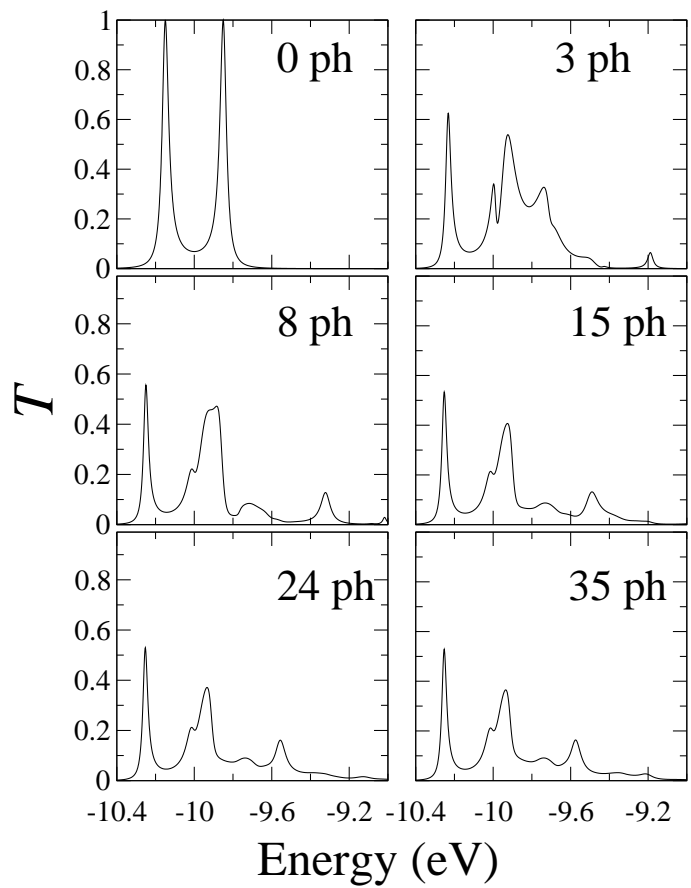}
\caption{\footnotesize{Total transmission coefficient
    (elastic plus inelastic) as a function of the energy obtained for an increasing number of
    phonon states $N_\mathrm{ph}$. The convergence is obtained for 24 states (solid line), that is a
    maximum occupation of $n_\alpha$=5 for each phonon mode.}}
\label{displ0tT}
\end{figure}

In figure \ref{displ0tT} we show the total transmission coefficient (elastic plus inelastic) 
at zero bias as a function of energy for different numbers 
of phonon states $\ket{n_1,n_2}$ included in the calculation. 
It is clear that as $N_\mathrm{ph}$ increases the transmission coefficient 
becomes more broadly distributed in energy and the resonances are largely 
smeared out. When no phonons are present $T(E)$ has only two sharp 
resonances, located at the bonding and anti-bonding energies of the molecule,
with a broadening proportional to the electronic coupling with the
leads. The inclusion of phonon absorption/emission opens new scattering 
channel where phonon-assisted hopping to the molecular level becomes 
possible. This broadens the spectral range of $T(E)$ and shifts the 
transmission peaks. In particular the resonance at the bonding state 
shifts almost rigidly of about 0.15~eV towards lower energies, while
that at the antibonding level gets substantially smeared. 

This behavior is somehow expected since in our working conditions of low temperature
electrons cannot be transmitted with a net energy gain (no net phonon absorption
is allowed). Thus most of the spectral modifications due to electron-phonon 
scattering occur above the Fermi level. If we now focus our attention on the
transmission coefficient for $N_\mathrm{ph}$=24 (the higher occupation allowed
is $n_\alpha$=5 for both the phonon modes) we can associate the first peak at about
-10.25~eV with resonant elastic transport through the bonding state, the second
just below -10.0~eV with phonon-assisted resonant transport through the bonding state
via the first phonon mode ($\hbar\omega_1$=0.24~eV), the third at about -9.92~eV
with phonon-assisted resonant transport through the bonding state
via the second phonon mode ($\hbar\omega_2$=0.33~eV), and so on.
Note that peaks in the transmission coefficient appearing at rather high energies
are the result of resonant transport through either the bonding or the antibonding
level via multiple phonon emission.

Note also that upon increasing the number of phonon states $T(E)$ is only weakly affected
for low energies, but changes drastically at high energies. This is because new scattering
paths involving multiple phonon emission become available. However these high order
multiple scattering events are increasingly improbable and one expects $T(E)$
to saturate for $N_\mathrm{ph}$ large enough. As a measure of the convergence 
of $T$ with the number of phonon modes we evaluate the mean deviation $\Delta_{mn}$
between the transmission coefficients $T_m$ and $T_n$ calculated respectively for
$N_\mathrm{ph}$=$m$ and $N_\mathrm{ph}$=$n$
\begin{equation}
\Delta_{mn}=\frac{1}{M}\sum_i^{M}|T_m(E_i)-T_n(E_i)|\;,
\end{equation}
with the sum running over the energy mesh points. The values of $\Delta_{mn}$ for the 
transmission coefficients of figure \ref{displ0tT} are respectively $\Delta_{0,3}$=0.0242, 
$\Delta_{3,8}$=0.0121, $\Delta_{8,15}$=0.0065, $\Delta_{15,24}$=0.0030 and 
$\Delta_{24,35}$=0.0012. A value of 
$\Delta\sim 0.0015$ is usually considered as a good level of convergence.

\begin{figure}[htbp]
\centering
\includegraphics[width=8truecm]{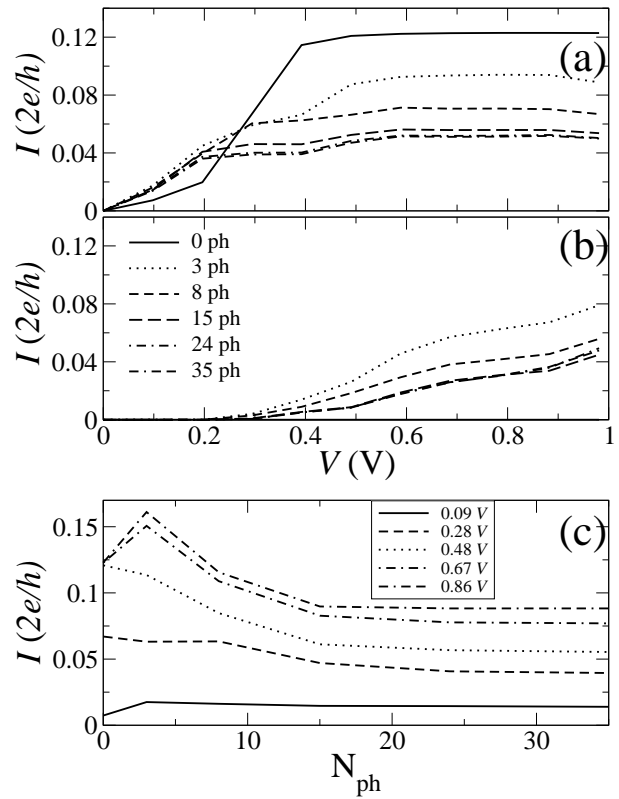}
\caption{\footnotesize{a) Elastic current for different phonon cutoffs. b)
Inelastic current for different phonon cutoffs. c) Total current
(elastic+inelastic) as a function of number of phonons. Each line
represents different voltages. At V$=0$ the total current is always 0.}}
\label{current_ela}
\end{figure}

The convergence of the transmission coefficients clearly guarantees the convergence 
of all the quantities related to them. Therefore after having determined $N_\mathrm{ph}$ 
we can then calculate the current using equation (\ref{currentf}). In figure
\ref{current_ela} we show the elastic and inelastic contributions to the
$I$-$V$ characteristic and the total current (elastic plus inelastic) as
a function of $N_\mathrm{ph}$ for different biases. A simple rule of
thumbs for understanding the features of the $I$-$V$ curve is that most of the current originates
from the total transmission coefficient in an energy window comprised between 
$E_\mathrm{F}+V/2$ and $E_\mathrm{F}-V/2$. Note that this is only approximately valid 
when electron-phonon scattering is present. In fact the self-consistent electron distributions 
$f$ can provide some non-vanishing contributions to the current coming from 
transmission coefficients outside this energy window.

First consider the case in which $N_\mathrm{ph}$=0, i.e. the purely elastic
case. The $I$-$V$ has only the elastic component and shows three different
slopes, corresponding to three different resistances. For bias below 0.2~Volt
the transmission is small, since no molecular state is comprised in the bias window. 
Then for voltages between 0.2 and 0.4~Volt, both the bonding and the antibonding 
states of the molecule appear in the bias window, producing a large increase of the
current. Finally for higher biases the current saturates since no new molecular states
are available for transport. 

The main effects of switching on the electron-phonon interaction $N_\mathrm{ph}\ne0$
is that the inelastic current starts to participate to the conductance. This is almost zero
for biases up to about 0.2~Volt, which is roughly the energy of the first phonon mode,
and then increase almost linearly with bias. This opening of phonon-assisted transport channels
appears in the $I$-$V$ curve as a peak in the derivative of the differential conductance,
a property which is used for detecting molecular vibrations from transport measurements
\cite{iest,smit,rotational}. Note that phonon emission for biases smaller
than the energy of the first phonon mode is certainly possible. However this process contributes
little to the current, since the final state has high probability to be filled. Note also that the elastic current
for small biases increases with respect to the purely elastic case. This is due to the contribution
of multiple scattering, which effectively increases the value of the transmission coefficient around
$E_\mathrm{F}$ (see figure \ref{displ0tT}). The relative portions of elastic and inelastic current
depend on the details of the Hamiltonian, however our analysis demonstrate that a considerable
number of phonon states must be included in the calculations for saturating the current.

\begin{figure}[htbp]
\centering
\includegraphics[width=8truecm]{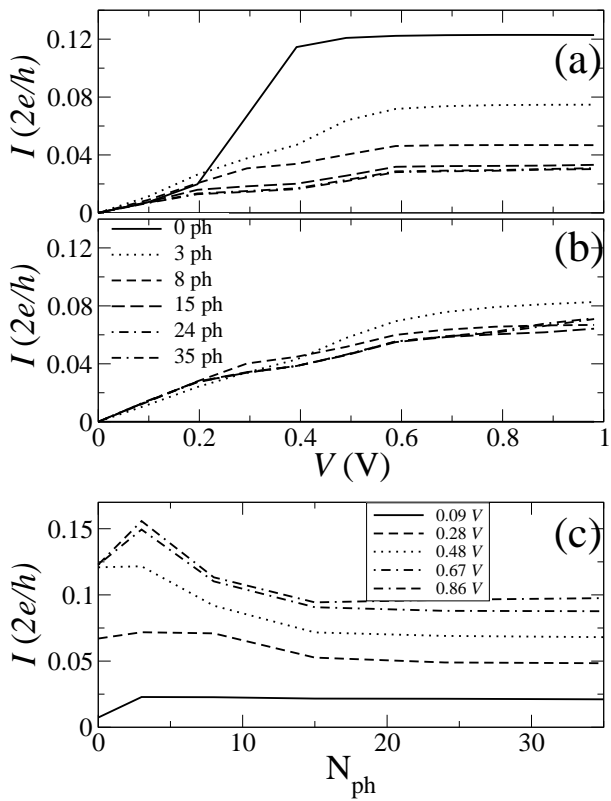}
\caption{\footnotesize{a) Elastic current for different phonon cutoffs,
calculated without considering the exclusion principle. b)
Inelastic current for different phonon cutoffs, calculated without considering
the exclusion principle. c) Total current(elastic+inelastic) calculated without considering the
exclusion principle
(elastic+inelastic) as a function of the number of phonons. Each line
represents different voltages. At V$=0$ the total current is always 0.}}
\label{displ0tC}
\end{figure}

Figure \ref{displ0tC} shows the same currents of figure \ref{current_ela}, this time obtained without
imposing the Pauli exclusion principle through the distributions $f$. 
The most evident difference is that now the inelastic current does not vanish for 
$V<0.2$~Volt, i.e. for biases smaller than the energy of the first phonon mode. 
This happens since the occupation of the final state is not considered, thus transmission
to final energies below $E_\mathrm{F}-V/2$ (the chemical potential of the right-hand 
side lead) is possible. As a consequence the elastic current in the same bias region is reduced
with respect to the previous case. Finally note that the currents calculated either imposing or
not imposing the Pauli principle are similar for large biases, where the vast majority of the
incoming states is filled, and the final states are empty.

%%%%%%%%%%%%%%%%%%%%%%%%%%%%%%%%%%%%%%%

\subsection{Transport through H$_2$ molecules}

Platinum point contacts sandwiching H$_2$ molecules represent the
ideal system, where to investigate the effects of electron-phonon interaction on the electron 
transport. These contacts in fact are stable for long times, enabling the measurement
of low bias $I$-$V$ characteristics under different stretching conditions \cite{smit}. 
The most important aspect of the current/voltage curves is a rather sharp drop of
the differential conductance $G_\mathrm{R}(V)=\frac{d I}{d V}$ from a value of approximately 
$G_0=2e^2/h$ for voltages close to 63~mVolt. This corresponds to a peak in the 
derivative of $G_\mathrm{R}$, $\frac{d G_\mathrm{R}}{d V}$, and it 
is attributed to the onset of electron-phonon 
scattering to the lowest phonon mode available in the junction \cite{smit}. 

Despite the initial controversy \cite{Gar04} recent {\it ab initio} calculations 
\cite{VictorH2,H2stretch} 
have convincingly proved that the transport in this system occurs through 
the antibonding state of the H$_2$ molecule in the so called ``bridge'' configuration, 
i.e. with the molecular bond lying parallel to the transport direction. The
antibonding state provides a high transmission channel and the differential
conductance is about 0.9~$G_0$. Moreover {\it ab initio} calculations have also
suggested that a transversal vibrational mode of the H$_2$ molecule is
responsible for the differential conductance drop at $\sim$60~meV  \cite{H2stretch}.
This last identification, although supported by the blue-shift of the differential
conductance drop upon stretching, however is more difficult to establish since the local
arrangement of the Pt atoms at the tip is unknown (it is likely that the H$_2$ molecule
bridges the Pt contacts in a zig-zag geometry \cite{VictorH2}). 

In an attempt of understanding the main mechanism responsible for the
drop of $G_\mathrm{R}(V)$ and to provide a further test to our scheme, 
we model a Pt-H$_2$-Pt point contact as two semi-infinite monoatomic 
current/voltage leads sandwiching a H$_2$ molecule. The leads are at 
half-filling with on-site energy $\epsilon_0$=0~eV and hopping parameter 
$\gamma$=5.0~eV. This guarantees a rather large (20~eV) bandwidth. 
The tight-binding parameters
for the H$_2$ molecule are chosen in such a way of aligning the antibonding state 
at the Fermi level and to give a conductance close to 0.97~$G_0$ for zero bias
\cite{smit}. We consider as relevant phonon modes only longitudinal modes and we 
tune the electron-phonon coupling in such a way of reproducing the experimental drop 
of the differential conductance. Note that at this level of approximation the choice 
of using either transverse or longitudinal modes is somehow immaterial.
\begin{figure}[htbp]
\centering
\includegraphics[width=7truecm]{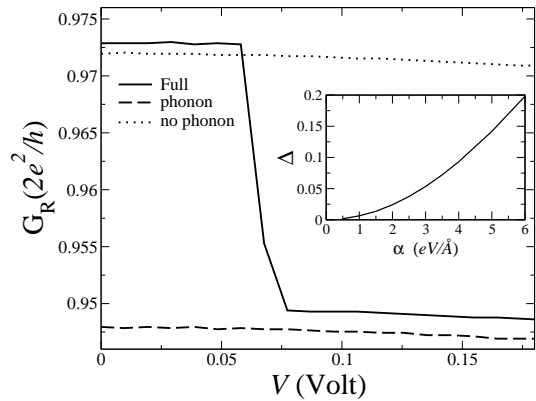}
\caption{\footnotesize{Differential conductance for the H$_2$ molecule sandwiched
between Pt leads. The solid line is for the full calculation where both the Pauli principle and 
electron-phonon interaction are included, the line labeled with ``phonons'' is for a 
the calculation done without considering the Pauli principle and the line labeled with 
``no phonons'' is for purely elastic transport. In the inset the drop of the differential 
conductance as a function of the electron-phonon coupling $\alpha$. The relative drop 
$\Delta$ is defined as the ratio between the differential conductance at $V$ just above 
the first phonon energy ($V$=63~mV) and that at zero bias 
$\Delta=[G_\mathrm{R}(V)-G_\mathrm{R}(0)]/G_\mathrm{R}(0)$.}\label{H2G}}
\end{figure}

Figure \ref{H2G} summarizes the main result of our calculations. Clearly, when both
electron-phonon interaction and Pauli principle are considered the differential 
conductance has a drop for voltages around 60~mV, which corresponds to the energy of the first
longitudinal phonon mode at 63~meV (oscillation of the center of mass of H$_2$). 
This drop of $G_\mathrm{R}$ is due to the enhancement of 
inelastic back-scattering, which reduces the transmission of the
elastic channel resonant at the antibonding state and it has the net effect of suppressing the
current. In fact a careful analysis of the electron distributions $f$ has revealed 
a peak in $f_{-,in}^L$ for energies corresponding to the emission of the first longitudinal
phonon. This emission, leading to the enhanced inelastic back-scattering is Pauli
suppressed for biases $V$ such that $eV<\hbar\omega_1$, but it becomes possible as soon
as the bias window is large enough to allow the phonon emission. Furthermore, within our
simplified model we do not expect any additional differential conductance drop for small bias,
since the next longitudinal phonon mode has a rather high energy ($\sim$430~meV).

Note that such a drop of $G_\mathrm{R}$ for $eV\sim\hbar\omega_1$ cannot be described
unless both electron-phonon coupling and Pauli principle are considered. 
On the one hand, when the Pauli principle is neglected the inelastic back-scattering 
is always present and contribute to the current, with the net result that
no conductance drop is found. In this case the conductance has 
a value similar to that obtained for $V>$~63~mV. On the other hand, if electron-phonon interaction
is set to zero, no drop can be found since inelastic back-scattering is not included. The
conductance now is similar to that calculated for  $V<$~63~mV and almost independent
from the bias. 

Importantly the magnitude of the differential conductance drop depends 
on the strength of the electron-phonon coupling, and it grows as the 
coupling is enhanced. Such a behavior is demonstrated in the inset of figure 
where $[G_\mathrm{R}(V)-G_\mathrm{R}(0)]/G_\mathrm{R}(0)$
is plotted as a function of the coupling $\alpha$. Interestingly we found
that $\Delta\propto\alpha^2$. Since the relevant longitudinal phonon mode is 
that of the center of mass of H$_2$ oscillating between the two Pt contacts,
the only $\alpha$ determining the transport is that of the Pt-H bond. This
is mainly an ss$\sigma$ bond \cite{VictorH2}, which roughly scales
as $1/d^2$ with the separation $d$ between Pt and H \cite{Harrison}.
Such a scaling leads to $\Delta\propto 1/d^6$, and therefore we predict
for longitudinal modes a very drastic dependence of the differential conductance
drop as a function of the point contacts separation.

%%%%%%%%%%%%%%%%%%%%%%%%%%%%%%%%%%%%%

\subsection{Molecular Spin-valves}

\begin{figure}[htbp]
\centering
\includegraphics[width=8truecm]{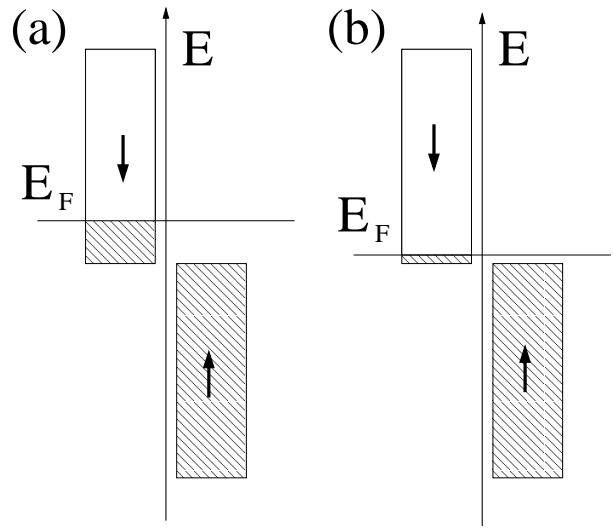}
\caption{\footnotesize{Schematic representation of density of states for the 
model $M_{\beta^{\prime}}$ half-metal and the two configurations studied: 
a) Fermi level well within the minority band, b) Fermi level
close to the minority lower band-edge.}}
\label{SysMagn}
\end{figure}

Finally we investigate the effects of inelastic electron-phonon scattering
on the transport properties and the GMR of a model spin-valve formed by half-metal 
current/voltage probes sandwiching an organic molecule. In particular
we consider a $M_{\beta^{\prime}}$ half-metal \cite{HM}, in which the Fermi 
level cuts through the minority band (see figure \ref{SysMagn}). In our model
calculation this is obtained by a linear chain of hydrogenic atoms whose on-site
energies are spin-split. In particular we consider $\epsilon^{\uparrow} = -14$ eV  
for the majority ($\uparrow$) and $\epsilon^{\downarrow} =-6$ eV for the minority 
($\downarrow$) spin bands. For both the spin-bands the hopping parameters are
$\beta_\mathrm{L}=\beta_\mathrm{R}=-2$ eV, which give a bandwidth 
and an exchange of 8~eV. With this choice of parameters the majority upper
band-edge coincide with the lower minority band-edge at $E$=-10~eV.

The molecules are linear chains of various lengths (from one to three atoms) 
made by atoms of mass 13~a.m.u. (C-H groups) and characterized by the hopping integral $t_{ij}=0.2$~eV. 
The on-site energy of the molecule is always close to the position of the Fermi level 
giving resonant transport through the molecular states. We have then
investigated two possible situations. In the first one (see figure \ref{SysMagn}a)
$\epsilon_\mathrm{M}=-8.2$~eV and the Fermi level 
$E_\mathrm{F}=-8.0$~eV is well within the minority band. 
In the second one $E_\mathrm{F}=\epsilon_\mathrm{M}=-9.7$~eV is close to 
the lower minority band edge (figure \ref{SysMagn}b).
The hopping parameters between the molecule and the leads 
and the electron-phonon couplings are the same as reference \cite{Eldon}, 
i.e. $w^0$=0.15~eV, $\alpha= 6.1$~eV/\AA, $K_\mathrm{lead-M}=180$~eV/\AA$^2$\ 
and $K_\mathrm{M}=90$~eV/\AA$^2$. This leads to phonon modes with the following 
energies: i) one atom chain, $\hbar\omega_1$=0.34~eV; ii) two-atom chain 
$\hbar\omega_1$=0.24~eV, $\hbar\omega_2$=0.33~eV; iii) three-atom
chain $\hbar\omega_1$=0.17~eV, $\hbar\omega_2$=0.30~eV
and $\hbar\omega_3$=0.34~eV.

In investigating the spin-transport we consider the two spin fluids approximation
where spin-flip scattering is neglected and the two spin sub-bands conduct in
parallel. This is a good approximation for C-based organic molecules since
both spin-orbit and hyperfine interactions are weak. The GMR is then defined as
usual as
\begin{equation}
MR=\frac{I_\mathrm{P}-I_\mathrm{AP}}{I_\mathrm{P}+I_\mathrm{AP}}\;,
\label{magnres}
\end{equation} 
where $I_\mathrm{P}$ is the current of the spin-valve when the magnetization vectors
of the two leads are parallel to each other and $I_\mathrm{AP}$ that for the antiparallel
alignment. Within the two spin fluid model these read $I_\mathrm{P}=
I^{\uparrow\uparrow}+I^{\downarrow\downarrow}$ and $I_\mathrm{AP}=
I^{\uparrow\downarrow}+I^{\downarrow\uparrow}$, with $I^{\alpha\beta}$
the spin current and the index $\alpha$ ($\beta$) indicating the spin direction
in the left-hand side (right-hand side) lead. Thus for instance $I^{\uparrow\downarrow}$
is the spin current for the antiparallel configuration where electrons propagate in
the majority band in the left-hand side lead and in the minority in the right-hand
side lead. 

Let us consider first the case when 
$\epsilon_\mathrm{M}=-8.2$~eV and $E_\mathrm{F}=-8.0$~eV.
In this situation the current is not negligible only for the minority spins
in the parallel configuration $I^{\downarrow\downarrow}$ at least for biases 
lower than 3.6~Volt, after which also $I^{\uparrow\downarrow}$ gives contributions.
In fact the particular density of states of our model half-metal and the 
relatively large Fermi level, result in the fact that for all the other configurations  
the density of states (DOS) of one of the two leads vanishes within the relevant bias
window. The current is therefore completely suppressed by the lack of available states
in the leads, and the GMR is always 1. Note that in principle multiple phonon emission 
might result in finite current also for other configurations, however this is
always entirely suppressed by the Pauli exclusion principle.
\begin{figure}[htbp]
\centering
\includegraphics[width=7truecm]{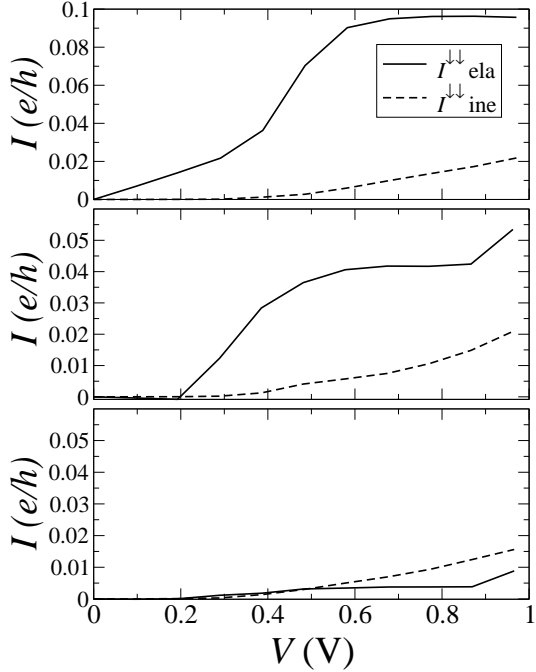}
\caption{\footnotesize{$I$-$V$ characteristic for the model spin-valve with 
$E_\mathrm{F}=\epsilon_\mathrm{M}=-8.2$~eV described in the text. The molecules
are formed respectively from one atom (upper panel), two atoms (middle panel) and
three atoms (lower panel). The only non-negligable contribution to the current is
from the minority electrons in the parallel configuration $I^{\downarrow\downarrow}$.}}
\label{MagnRes82}
\end{figure}

The calculated $I$-$V$ for the parallel case (up to 1 Volt) is presented 
in figure \ref{MagnRes82} for the three atomic chains investigated. The most notable 
feature is the drastic suppression of the elastic current as the chain length increases,
and the fact that the inelastic component is only weakly affected. This is
somehow expected since as the molecules becomes longer the direct
tunneling probability decays and the current is dominated by the 
inelastic component. Interestingly, for this choice of parameters,
reproducing (CH)$_n$ molecules, the inelastic current dominates already for
rather short lengths  ($n$=3).

The second case, when $E_\mathrm{F}=\epsilon_\mathrm{M}=-9.7$~eV is somehow
different, since the Fermi level cuts much closer to the minority lower band-edge. In this
situation the half-metal DOS suppresses the current only at low bias and one expects
transport in the antiparallel configuration, due to the $I^{\uparrow\downarrow}$ channel
for biases as small as 0.6~eV. In fact at such biases finite DOS appears in the bias
window for both the leads. This does not happen for the $I^{\uparrow\uparrow}$
and $I^{\downarrow\uparrow}$ contributions, which vanish at all biases. Therefore
one has $I_\mathrm{P}=I^{\downarrow\downarrow}$ and $I_\mathrm{AP}=I^{\uparrow\downarrow}$
and the GMR is in general dependent on the bias.
\begin{figure}[htbp]
\centering
\includegraphics[width=7truecm]{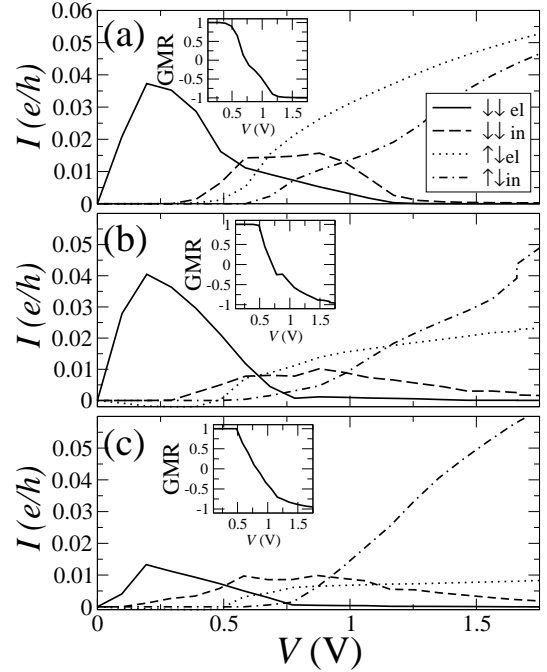}
\caption{\footnotesize{$I$-$V$ curves for the parallel and antiparallel configurations
of a half-metal molecular spin-valve made from atomic chains containing (a) one, 
(b) two and (c) three atoms. In this case $E_\mathrm{F}=\epsilon_\mathrm{M}=-9.7$~eV,
$I_\mathrm{P}=I^{\downarrow\downarrow}$ and $I_\mathrm{AP}=
I^{\uparrow\downarrow}$. In the insets the corresponding GMR as a function
of bias.}}
\label{MagnRes95}
\end{figure}

Figure \ref{MagnRes95} shows the $I$-$V$ curves for both the parallel and antiparallel 
configurations of the spin-valve and the corresponding GMR, for molecules made respectively
from one, two and three atoms. Let us consider first $I_\mathrm{P}=I^{\downarrow\downarrow}$.
The elastic component of the current dominates for voltages up to approximately 1.5~Volt,
after which it gets completely suppressed. This suppression originates from the fact that the 
``effective'' bias window (where both the leads have finite DOS) remains constant 
for $V>0.6$~Volt, but it moves increasingly far from the energy levels of the molecular
states. This drastically reduces the elastic transmission coefficient with a strong
decrease of the current. The same effects somehow occurs for the inelastic component of the
current although phonon emission allows transmission over a much wider energy 
window. This makes the inelastic component dominating for $V>1$~Volt, regardless of
the chain length.

In contrast in the antiparallel configuration no current can flow until the bias is large enough 
that the DOS of the two leads overlap for some energies. This happens for $V\sim$0.5~Volt
for the elastic component and $V\sim$1.0~Volt for the inelastic one. In this case we do not expect
a decrease in the current for increasing bias, since the bias window is not cut by any
band edge. As in the previous case the inelastic component becomes progressively more
dominant as the chains gets longer and it is expected to dominate completely for very long
chains.  The resulting GMR is strongly bias dependent. In particular, for the present choice of 
parameters, it changes sign from positive at small bias to negative a high bias.

\begin{figure}[htbp]
\centering
\includegraphics[width=7truecm]{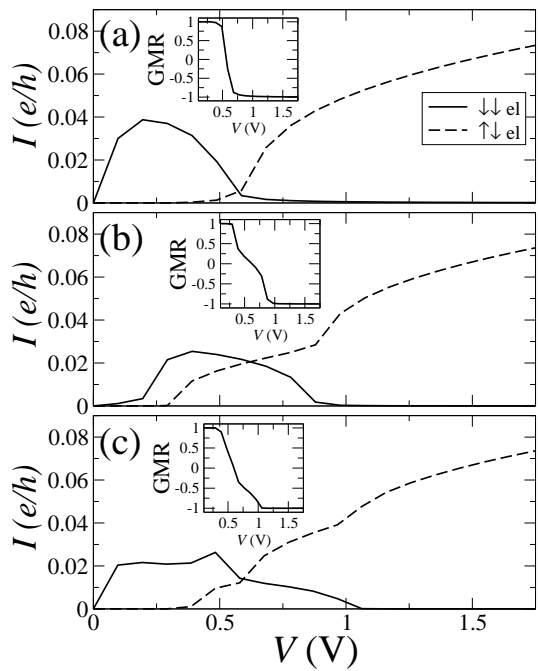}
\caption{\footnotesize{Elastic $I$-$V$ curves for the parallel and antiparallel configurations
of a half-metal molecular spin-valve made from atomic chains containing (a) one, 
(b) two and (c) three atoms. In these calculations no electron-phonon interaction is
considered. $E_\mathrm{F}=\epsilon_\mathrm{M}=-9.7$~eV
and $I_\mathrm{P}=I^{\downarrow\downarrow}$ and $I_\mathrm{AP}=
I^{\uparrow\downarrow}$. In the insets the corresponding GMR as a function
of bias.}}
\label{MagnResEl}
\end{figure}

As a comparison in figure (\ref{MagnResEl}) we present results for the same systems
obtained without electron-phonon interaction. In this case there is no sustantial 
reduction of the elastic current as the molecular chain becomes longer, and the 
transport is always resonant through the molecular levels. However in absence of
the inelastic component, the total current for the parallel alignment sharply
decays for biases in excess of 1~Volt. As a consequence the GMR sharply changes
from 1 to -1 as the bias increases. 

Generally, inelastic scattering increases
the spectral range (the energy region where the total transmission coefficient does
not vanish) contributing to the current (see figure \ref{displ0tT}). 
In this case of half-metallic leads, the effect is to change the GMR as a 
function of bias. However if the leads are made from transition metals, then 
the effect of electron-phonon scattering is expected to be a change in the 
spin-polarization of the current \cite{mazin}. This is connected to the increase of the 
spectral range of the transmission coefficient. A large
spectral range means that a large portion of the ferromagnetic DOS contributes to
the current. However in an ordinary transition metal the spin-polarization of the DOS
is large at around $E_\mathrm{F}$ but decreases for both higher and lower energies. 
At higher energies this is due to the fact that there is little contribution
from the $d$ electrons for both spin directions, while at lower energy a large
$d$-DOS is present for both spin directions. We therefore expect that electron-phonon
interaction will produce a general suppression of the GMR.

\section{Conclusions}

We have demonstrated the r\^ole of inelastic electron-phonon scattering on
the transport properties of different nanoscale devices. This is treated using
the multi-channel scheme introduced by Bon$\check{\mathrm{c}}$a and Trugman
with the explicit inclusion of the Pauli exclusion principle. Such method
allows us to map a scattering problem with inelastic effects, onto a fictitious
problem where only elastic scattering events are present. We have then given
a full description of the method, within our surface Green's function
formalism, and described the criterion of convergence over the number and
nature of phonon modes.

Two specific material systems have been investigated. The first are Pt
point contacts sandwiching a H$_2$ molecule. In this case the differential
conductance shows a sudden drop for a bias corresponding to
the energy of the first H$_2$ longitudinal phonon mode, which we have
explained as an enhancement of inelastic backscattering. Note that
such a drop cannot be found unless the Pauli principle is considered.

Then we have investigated two types of spin-valves, obtained by
contacting with half-metals a linear chain
of CH groups. In this case the GMR depends on the specific alignment of
the Fermi level with respect to the minority lower band-edge, and
in general it is strongly bias dependent. The presence of electron-phonon
scattering largely increases the current in the parallel state at high
voltages and makes the change in GMR upon bias smoother with
respect to the case where electron-phonons interaction is not considered.

\end{document}